\documentclass[prb, onecolumn, nofootinbib, citeautoscript, 10pt, longbibliography, notitlepage]{revtex4-2}

\usepackage{graphicx}
\usepackage{dcolumn}
\usepackage{bm}
\usepackage{amsmath}
\usepackage{tabularx,graphicx}
\usepackage{epstopdf}
\usepackage{latexsym}
\usepackage{amssymb}
\usepackage{color, colortbl}
\usepackage{psfrag}
\usepackage{bbm}
\usepackage{titlesec}
\usepackage{dsfont}
\usepackage{feynmp}
\usepackage{slashed}
\usepackage{multirow}
\usepackage{physics}
\usepackage[tight]{subfigure}

\usepackage[papersize={8.5in,11in}]{geometry}

\usepackage{color}
\definecolor{darkblue}{rgb}{0.,0.,0.4}
\definecolor{darkred}{rgb}{0.5,0.,0.}
\definecolor{BlueViolet}{RGB}{138,43,226}
\definecolor{SkyBlue}{RGB}{30,144,255}
\definecolor{DarkGreen}{RGB}{0,100,0}
\usepackage[pdftex,colorlinks=true,linkcolor=darkblue,citecolor=blue,urlcolor=darkred]{hyperref}

\def \nn{\nonumber \\}

\begin{document}

\title{Delta-function-potential junctions with quasiparticles occupying tilted bands with quadratic-in-momentum dispersion}

\author{Ipsita Mandal}
\email{ipsita.mandal@snu.edu.in}

\affiliation{Department of Physics, Shiv Nadar Institution of Eminence (SNIoE), Gautam Buddha Nagar, Uttar Pradesh 201314, India}

\begin{abstract}
We continue our explorations of the transport characteristics in junction-configurations comprising semimetals with quadratic band-crossings, observed in the bandstructures of both two- and three-dimensional materials. Here, we consider short potential barriers/wells modelled by delta-functions. We also generalize our analysis by incorporating tilts in the dispersion. Due to the parabolic nature of the spectra, caused by quadratic-in-momentum dependence, there exist evanescent waves, which decay exponentially as we move away from the junction represented by the location of the delta-function potential. Investigating the possibility of the appearance of bound states, we find that their energies appear as pairs of $\pm |E_b |$, reflecting the presence of the imaginary-valued wavevectors at both positive and negative values of energies of the propagating quasiparticles.
\end{abstract}
\maketitle

\tableofcontents

\section{Introduction}

\begin{figure*}[t]
\centering
\subfigure []{\includegraphics[width=0.25 \textwidth ]{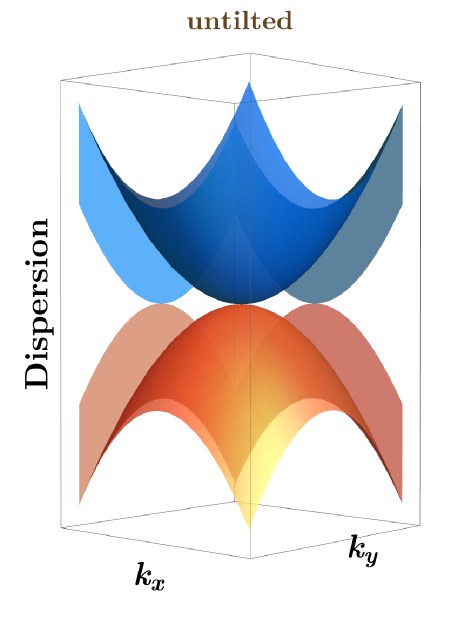}} \qquad
\subfigure []{\includegraphics[width=0.25 \textwidth ]{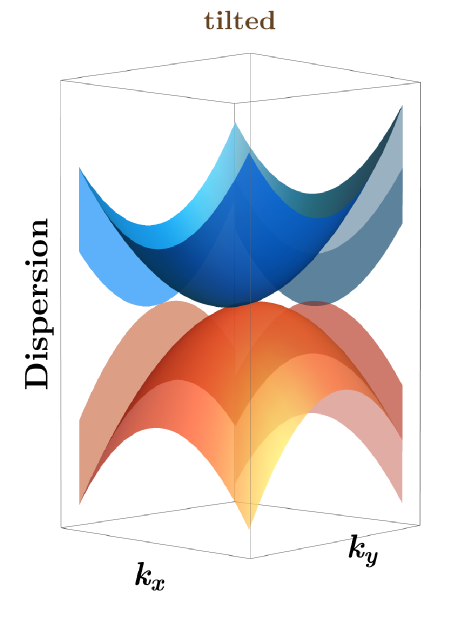}} 
\caption{Schematic dispersion of a QBCP against the $k_x k_y$-plane for (a) zero tilt and (b) nonzero tilt. The 
tilting is with respect to the $k_y$-direction, which is depicted with greater clarity through the the projections of the dispersion along the respective momentum axes.
\label{figdis}}
\end{figure*}

Twofold- and multifold-degenerate points, caused by two or more bands crossing at specific points in the Brillouin zone (BZ), lead to the emergence of systems called nodal-point semimetals. Consequently, the density-of-states goes exactly to zero right at the nodal points, and such a behaviour effectively falls in between that of metals (with bands overlapping in finite regions of the BZ) and insulators (with a finite gap between the bands everywhere). The nodal points, thus, represent Fermi points (rather than Fermi surfaces) when the chemical potential is adjusted to cut right at those points, and appear both in two-dimensional (2D) and three-dimensional (3D) materials. For example, graphene represents the 2D case when the honeycomb lattice is at half-filling, leading to 2D Dirac cones \cite{graphene_review} in the BZ. Considering 3D, we are well-acquainted with the Dirac and Weyl semimetals \cite{armitage_review}. While these well-known examples constitute the simplest cases of linear-in-momentum isotropic dispersion in the vicinity of the nodes, we are aware of bandstructures hosting anisotropic and/or nonlinear-in-momentum dispersion-directions. In particular, in this paper, we will focus on 2D and 3D semimetals hosting quadratic band-crossing points (QBCPs) \cite{kai-sun, tsai, ips-seb, Abrikosov, LABIrridate, MoonXuKimBalents, rahul-sid, ipsita-rahul, ips_tunnel_qbcp, armitage, ips-hermann1, ips-hermann2, ips-hermann3, ips_qbt_plasmons, ips-jing, ips-sandip, ips_tunnel_qbcp_corr}, where a possible anisotropy appears in the parabolic-energy spectrum via tilting of the dispersion (with respect to a given momentum-axis). \textcolor{black}{Schematically, the dispersion is shown in Fig.~\ref{figdis}, corresponding to the Hamiltonians in Eqs.~\eqref{eqham2D} and \eqref{eqham3D} shown later.} The 2D QBCPs are exemplified by bilayer graphene \cite{falko_ll, geim_ll, geim}, and lattice structures of the checkerboard \cite{kai-sun} (at half-filling), Kagome \cite{kai-sun} (at one-third-filling), and Lieb \cite{tsai} types. In 3D, materials like gapless semiconductors in the presence of a sufficiently strong spin-orbit coupling \cite{Beneslavski}, gray tin ($\alpha$-Sn), mercury telluride (HgTe), and pyrochlore iridates (with the chemical formula $\text{A}_2\text{Ir}_2\text{O}_7$, where A stands for a lanthanide element~\cite{pyro1,pyro2}) host parabolic spectra around nodal points. 
The 3D QBCPs have also been often dubbed as ``Luttinger semimetals'' in the literature, originating from the fact that their low-energy effective behaviour is captured by the so-called Luttinger Hamiltonian \cite{Abrikosov, luttinger, murakami, igor16}.

For the mesoscopic systems considered in this paper, the situation is described by the Landauer–Büttiker single-particle formalism for the conductance of nano-scale coherent systems, where ``coherence'' means that the quantum-mechanical coherence length is longer than the sample size. The standard set-up involves a mesoscopic sample connected to electron reservoirs (or contacts) in the form of macroscopic metal contacts. Basically, the coherent movement of the electrons through the sample is achieved by restricting the sample’s size to be smaller than the mean free path of an electron. The cleaner the sample, the longer the mean free path is and, thereby, the allowed system size increases. Since the reservoirs are macroscopic conductors, much larger than the consideration of the mesoscopic region, we can safely assume that the electrons entering the reservoir will be thermalized at the temperature and chemical potential of the contact before returning to the mesoscopic sample. The contact is thus required to be reflectionless, meaning that an electron impinging on a contact will be fully absorbed and thermalized by the contact before being re-emitted into the sample. This way of treating conductance in mesoscopic systems is often dubbed as the two-probe Landauer–Büttiker formalism. Now, the conductance in a two-terminal mesoscopic sample can be computed in two different ways (see chapter 7 of Ref.~\cite{bruus}, for example): (1) On physical grounds,the population of the scattering states is given by the equilibrium distribution functions of the reservoirs, which allows us to calculate the current directly. This leads to the celebrated Landauer formula, which tells us that the conductance of a mesoscopic sample is given by the sum of all the transmission possibilities that an electron has, when propagating with an energy equal to the chemical potential \cite{landauer, buttiker, blanter-buttiker}. The Landauer formula essentially shows that the conductance of a mesoscopic sample is quantized in units of $2 \,e^2 /h$. The number of quanta will be the number of transmission possibilities, denoted as channels, connecting the two ends of the mesoscopic sample.
(2) The conductance can be computed using linear response theory, which involves computing the the current-current correlation function to extract the conductivity.

In an earlier work \cite{ips_tunnel_qbcp_corr}, we have computed the transmission and reflection coefficients of the quasiparticles in the vicinity of 2D and 3D QBCPs, when moving across a rectangular potential barrier. Given the nonlinear-in momentum nature of the dispersion, there appear evanescent waves (see Refs.~\cite{deng2020, ips-aritra, banerjee, ips-abs-semid} for analogous situations in semimetals hosting hybrid dispersions depending on the momentum-axes considered). In this paper, we aim to chalk out analogous transport properties, but considering a delta-function-potential well/barrier. Additionally, we consider a tilt in the spectrum, which should exist in generic bandstructures. Plugging in the transmission probabilities ($T_{\mathbf n}$'s), the conductance ($G$) is given by the Landauer formula of \cite{landauer, buttiker, blanter-buttiker}
\begin{align}
\label{eqland}
G = \frac{e^2} { 2\, \pi\, \hbar} 
\sum_{\mathbf n}
 T_{\mathbf n} \,,
\end{align}
ignoring the spin-degeneracy.
Here, $e$ is the charge of one electron and $\mathbf n$ labels the transverse momentum modes in the the strip of the material (in the experimental set-up). When an external potential difference $\Phi $ is applied across a circuit, shot noise is the physical quantity which provides a measure of the fluctuations of the electric current-density away from its average value. At zero temperature, while the actual shot noise is defined by 
$ \mathcal S =  \frac{e^3\, |\Phi| } {  \pi\, \hbar} 
\sum_{\mathbf n}
 T_{\mathbf n} \left( 1 - T_{\mathbf n} \right ) ,$
the Poisson noise is given by the expression
$
\mathcal S_P =  \frac{e^3\, |\Phi| } {  \pi\, \hbar} 
\sum_{\mathbf n} T_{\mathbf n} $ \cite{blanter-buttiker}.
The Poisson noise is the value of the noise that would be measured if the system produced noise due to quasiparticles carrying a single set of the relevant quantum numbers. A convenient measure of the sub-Poissonian shot noise is the Fano factor ($F$), which is the ratio defined as \cite{blanter-buttiker}
\begin{align}
\label{eqfano}
F = \frac{\mathcal S}{ \mathcal S_P } =
\frac{\sum_{\mathbf n}
 T_{\mathbf n} \left( 1 - T_{\mathbf n} \right )}
 { \sum_{\mathbf {\tilde n}} T_{\mathbf{\tilde  n}} } \,.
\end{align}

The paper is organized as follows. Secs.~\ref{sec2Dmodel} and \ref{sec3Dmodel} deal with the 2D and 3D QBCPs, respectively. The representative characteristics of $T_{\mathbf n}$, $G$, and $F$ are computed and compared with those of the free-electron gases. In Sec.~\ref{secbs}, we venture into determining the energies of the bound states that may appear in our junction configurations.
Finally, we conclude with a summary and outlook in Sec.~\ref{secsum}. In all our expressions, we resort to using the natural units, which indicates that the reduced Planck's constant ($\hbar $), the speed of light ($c$), and the magnitude of a single electronic charge ($e$) are each set to unity.

\section{2D QBCP}
\label{sec2Dmodel}

A QBCP in a 2D system harbours a twofold band-crossing, represented by the low-energy effective Hamiltonian \cite{kai-sun},
\begin{align}
\tilde {\mathcal{H}}_{2D}^{\rm kin}(k_x,k_y)&=
 \frac{1 }{2\,m}
\left[ 2\, k_x\,k_y\, \sigma_x +  \left( k_y^2-k_x^2\right) \sigma_z  \right] 
+ \tilde \eta \, k_y\, \mathbb{I}_{2 \times 2} \,,
\end{align} 
in the momentum space. Here, the second term represents a tilting of the dispersion with respect to the $k_y$-momentum direction.
This Hamiltonian, with $\tilde \eta =0 $, also represents the chiral charge-carriers of bilayer-graphene \cite{falko_ll, geim_ll, geim}.

For the sake of uncluttering of notations, we multiply each momentum-component by the factor of $ \sqrt{2 \, m}$.
Hence, we deal with the scaled-momentum Hamiltonian captured by
\begin{align}
{\mathcal{H}}_{2D}^{\rm kin}(k_x,k_y)&=
  2\, k_x\,k_y\, \sigma_x +  \left( k_y^2-k_x^2\right) \sigma_z
+  \eta \, k_y\, \mathbb{I}_{2 \times 2}   \, ,
\label{eqham2D}
\end{align} 
where the tilt parameter $\eta= \sqrt{2\, m} \; \tilde \eta $.
The energy eigenvalues are captured by $ \eta \, k_y \pm \,\varepsilon_{2D}(k_x,k_y)$, where
\begin{align}
\varepsilon_{2D}(k_x,k_y) =     k_x^2 + k_y^2  \,.
\end{align}
Here, the ``$+$" and ``$-$" signs, in the usual conventions, refer to the positive-energy (or conduction) and negative-energy (or valence) bands, respectively.
The corresponding orthonormal eigenvectors are given by
\begin{align}
\Psi_+^T =\frac{1}{\sqrt{k_x^2+k_y^2}} \left( k_y  \quad  
k_x \right ) \text{ and }
  \Psi_-^T =\frac{1}{\sqrt{k_x^2+k_y^2}}  \left(  - \, k_x  \quad  
   k_y  \right )  ,
\end{align}
respectively.

The 2D system is modulated by a delta-function-like electric potential barrier/well,
\begin{align}
V ( x ) =  V_0 \,\delta(x)\,,
\label{eqpot}
\end{align}
positioned at $x=0$. Here, $ | V_0 | $ reflects the strength of the potential barrier. Depending on whether $V_0>0$ or $V_0 <0 $, we have either a barrier or a well. The total Hamiltonian for the entire configuration is captured by
\begin{align}
\mathcal{H}_{2D}^{\rm{tot}} &=
 \mathcal{H}_{2D}^{\rm kin}(- \, i \,\partial_x \,,  \, -\, i\,\partial_y)
 + V(x) \,,
\end{align} 
when expressed in the position space. Here, we will consider the transport of quasiparticles along the $x$-axis, across the delta-function potential. We fix the Fermi level at a value of $E >0$ in the region outside the potential barrier, which can be adjusted in an experimental set-up by either chemical doping or an external gate-voltage.

\subsection{Boundary conditions}

For a material of a sufficiently large transverse dimension $W$, the specific form(s) of the boundary conditions are irrelevant for the bulk response. We use this freedom to impose periodic boundary conditions along the $y$-direction. This implies that, for a physical wavefunction, $\Psi ^{\mathrm{{tot}}} (x,y)$, in the position space, we impose the condition
of $ \Psi^{\rm{tot}}(x,W) = \Psi^{\rm{tot}}(x,0) \,.$
The transverse momentum $k_y$ appears in the form of plane-wave solutions, leading to the ansatz for the $k_y$-dependent modes to behave as $\sim  e^{  i\,k_y \, y }$.
Therefore, for a mode with the transverse (to the propagation direction) momentum-component $ k_y $, we must have
$ |k_y | \leq  
\sqrt{ E +\eta^2 / 4} - \eta /4  $. The $k_x$-components of the wavevectors of the incoming, reflected, and transmitted waves are obtained by solving $
E^2 = \left [ k_x^2 + k_y \left( k_y +\eta \right) \right ]^2 \Rightarrow k_x^2
 = \pm \,  E - k_y \left(k_y+\eta \right) \,.$
Consequently, in the regions $x<0$ and $x>0$, this relation leads to the following four solutions: 
\begin{align}
\label{eqksol}
k_x = \pm \,\sqrt{  E - k_y \left(k_y+\eta \right)} \text{ and }
k_x = \pm \, i\, \sqrt{  E + k_y \left(k_y+\eta \right) } \,.
\end{align}
Due to the imaginary solutions, in addition to the propagating plane waves, there exist evanescent waves~\cite{banerjee,  deng2020, ips-aritra, soori, ips-abs-semid} characterizing exponentially decaying amplitudes (since solutions with exponentially increasing amplitudes are physically inadmissible).

We follow the usual procedure of matching the piecewise-continuous regions (see, for example, Refs.~\cite{salehi,beenakker}) to compute the reflection and transport coefficients. Here, we consider the transport of positive energy states (i.e., $\Psi_+$) corresponding to electron-like particles. The transport of hole-like excitations (i.e., $\Psi_-$) will be similar. Hence, the Fermi level outside the potential barrier is adjusted to the value $E $ (with $E>0$).
A mode $\sim  \Psi_{k_y} (x) \, e^{i \,k_y \,y}$ is constructed in a piecewise fashion as
\begin{align}
& \Psi_{k_y} (x)  =  \begin{cases} \phi_L (x)  & \text{ for } x \leq 0   \\
\phi_R (x)  &  \text{ for } x > 0
\end{cases} ,
\end{align}
where
\begin{align}
 & \phi_L (x)   =  \frac{ 
 \Psi_+ (  k_{\rm in},  k_y) \,  e^{ i\, k_{\rm in} \, x }
+  r \, \Psi_+ (  - \, k_{\rm in},  k_y) \,   e^{- i\, k_{\rm in} \, x }
} 
{\sqrt{ \mathcal{V}  }}
+
\tilde r \, \Psi_- (  - \,i \, \kappa ,  k_y) \,   e^{ - \kappa \,|x| } \, ,\nn 
& \phi_R (x)  = 
  t\, \Psi_+ ( k_{\rm in},  k_y)\,
 \frac{  e^{ i\, k_{\rm in} \, x }} 
{\sqrt{ \mathcal{V}  }}
+ \,\tilde t \, \Psi_- ( i \, \kappa ,  k_y) \,   e^{- \,\kappa \, x } \,,\nn 
& k_{{\rm in}}  = \sqrt{  E - k_y \left(k_y+\eta \right) }\,, \quad
\mathcal{V}  \equiv   
 |\partial_{k_{\rm in}} \varepsilon_{2D} (k_{\rm in}, k_y)|
=  2\, k_{\rm in}  \,, \quad
\kappa = \sqrt{ E + k_y \left(k_y+\eta \right)}\,.
\end{align}
Here, the magnitude of the group-velocity, captured by
$ \mathcal{V} $, is needed to define the scattering matrix comprising the
transmission ($t$) and reflection ($r$) amplitudes.

The boundary conditions can be obtained by integrating the equation, $\mathcal{H}_{2D}^{\rm{tot}}\, \Psi^{\rm{tot}} = E \, \Psi^{\rm{tot}} $, over a small interval around $x=0$, in two consecutive steps. This amounts to imposing the continuity of the wavefunction and a constraint on its first-order derivatives (with respect to $x$), as shown below:
\begin{align}
\label{eqbdy1}
& \phi_L (0) = \phi_R (0) \text{ and } 
\partial_x \phi_R  (x) \big \vert_{x=0}
- \partial_x \phi_L  (x) \big \vert_{x=0} =   V_0 \,  \phi_L  (0)\,. 
\end{align}
These conditions are sufficient to guarantee the continuity of the flux of the probability-current density along the $x$-direction. From the matching of the wavefunction and its derivatives, we have two matrix-equations from the two boundary conditions.
For 2D QPCBs, each of these matrix-equations can be separated into two components, since each wavevector has two components. Therefore, we have $  2\times 2 = 4 $ equations for the four undetermined coefficients
$\lbrace r, \, \tilde r,  \, t, \tilde t  \rbrace $. 

\subsection{Transmission coefficients, conductance, and Fano factors}

\begin{figure}[h!]
\subfigure{\includegraphics[width =  \textwidth]{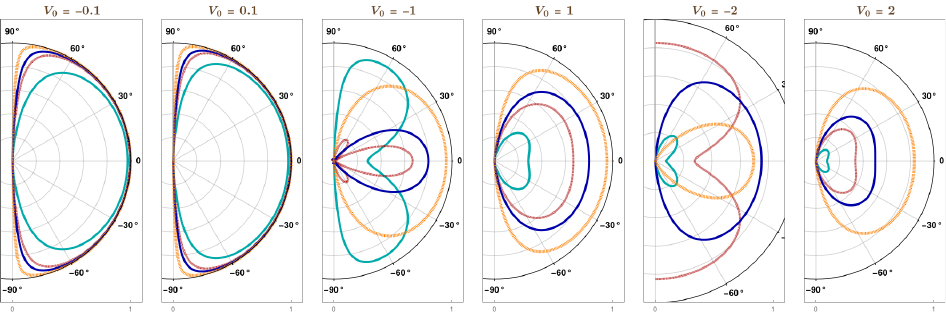}} 
\subfigure{\includegraphics[width =  \textwidth]{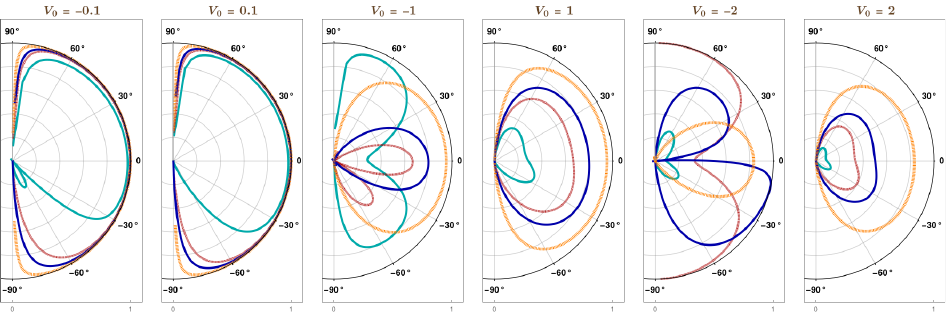}} 
\subfigure{\includegraphics[width =  \textwidth]{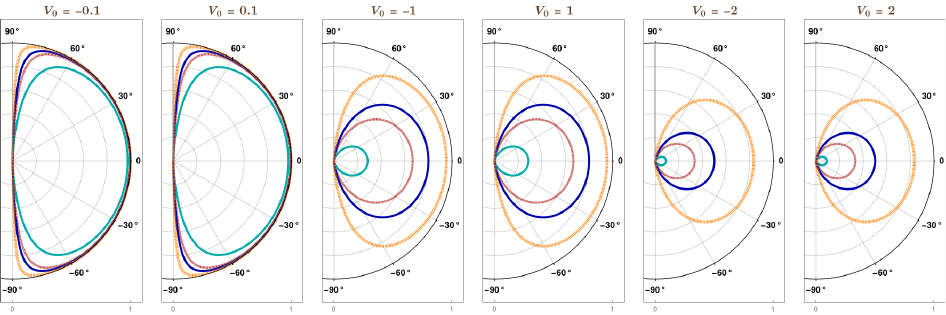}}\\
\subfigure{\includegraphics[width = 0.5 \textwidth]{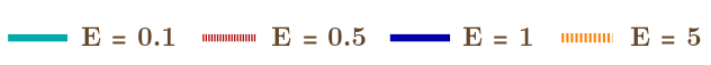}}
\caption{Plots of the transmission coefficient ($T$) as functions of the incident angle $\phi$, for various values of $ E $ and $V_0$, using Eqs~\eqref{eqT2D2} and \eqref{eqtval}. The subfigures in the top, middle, and bottom panels correspond to untilted 2D QBCP, tilted 2D QBCP with $\eta = 0.25 $, and 2D electron-gas, respectively.}
\label{figtrs2D}
\end{figure}

\begin{figure}[t]
\subfigure{\includegraphics[width = 0.75 \textwidth]{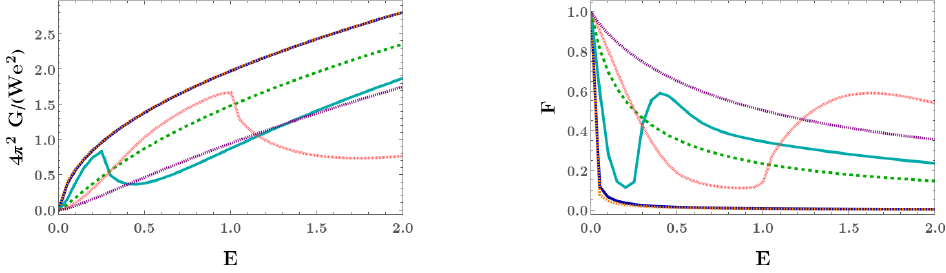}}
\subfigure{\includegraphics[width = 0.75 \textwidth]{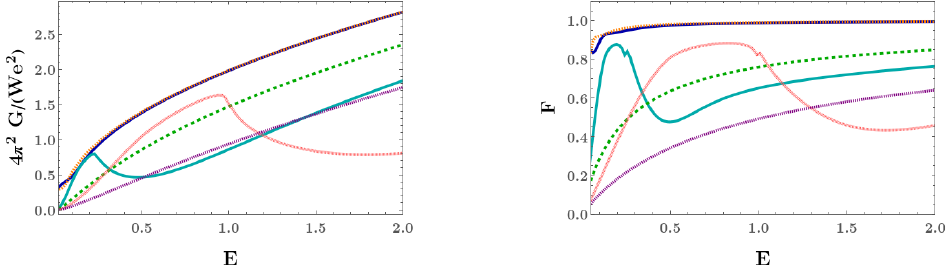}}
\subfigure{\includegraphics[width = 0.75 \textwidth]{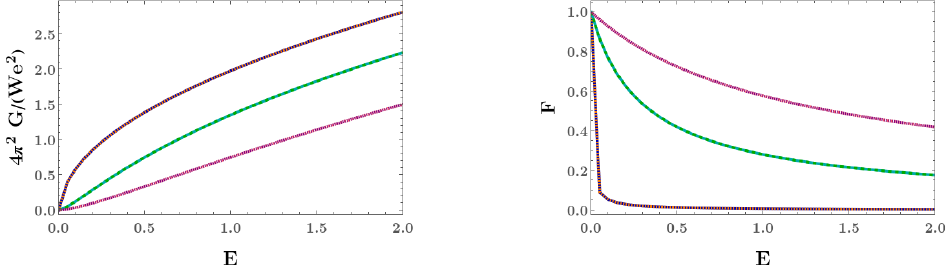}}
\subfigure{\includegraphics[width = 0.7 \textwidth]{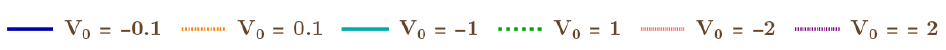}}
\caption{Plots of the conductance ($ G $) and Fano factor ($F$), as functions of $E $, for various values of $ V_0$. The subfigures in the top, middle, and lowest correspond to untilted 2D QBCP, tilted 2D QBCP with $\eta = 0.25 $, and 2D electron-gas, respectively.}
\label{figfano2D}
\end{figure}

The explicit analytical expressions for $t$ and $r$ are shown below:
\begin{align}
\label{eqtrs2D}
t &=
2 \, k_{in} \times \frac {
  \kappa^2 \, k_{in}^2
     \left (2 \, \kappa + V_ 0 \right)
   + 2 \, \kappa \, k_y^2
    \left[ \kappa\left (\kappa + V_ 0 \right) - k_{in}^2 \right ]
   - \left[k_y^4\left (2 \, \kappa + V_ 0 \right) \right ]}
{\left[\kappa  \, k_{in}\left (2 \, \kappa + V_ 0 \right)
    - 2 \, k_{in} \, k_y^2 - i \, V_ 0 \, k_y^2 \right ]
  \left[ k_y^2 \left (2 \, \kappa + V_ 0 \right)
+ \kappa  \, 
   k_{in}\left (2 \, k_{in} + i \, V_ 0 \right) \right ]}  \,,\nn
r & = \frac{ i\, V_ 0 \left (2 \, \kappa + V_ 0 \right)
\left (\kappa^2 \, k_{\rm in}^2 + k_y^4 \right)
}
{\left[\kappa  \, k_{in}\left (2 \, \kappa + V_ 0 \right)
    - 2 \, k_{in} \, k_y^2 - i \, V_ 0 \, k_y^2 \right ]
  \left[ k_y^2 \left (2 \, \kappa + V_ 0 \right)
+ \kappa  \, 
   k_{in}\left (2 \, k_{in} + i \, V_ 0 \right) \right ]} \,.
\end{align}
For the untilted case (i.e., $\eta = 0 $), we get
\begin{align}
\label{eqtrs2D2}
t &=
\frac {4 \, E  \cos \phi\, \sqrt {\sin^2\phi + 1} 
   - \sqrt {E} \, V_ 0\left   [ \cos  (3\phi) - 3 \cos  \phi \right  ]}
{\left[V_ 0 \, \sin^2 \phi + \sqrt {\sin^2 \phi + 1}
      \left (2 \, \sqrt {E} + i \, V_ 0 \cos \phi \right)
     \right]
  \left[\cos \phi
      \left (2 \, \sqrt {E}
       + V_ 0\sqrt {\sin^2 \phi + 1} \right)
    - i\, V_ 0 \sin^2 \phi \right ]}\,,\nn
r & = \frac{ i\, V_ 0 \left (2 \, \sqrt {E}
    \, \sqrt {\sin^2 \phi + 1} \, + V_ 0 \right)
}
{\left[V_ 0 \, \sin^2 \phi + \sqrt {\sin^2 \phi + 1}
      \left (2 \, \sqrt {E} + i \, V_ 0 \cos \phi \right)
     \right]
  \left[\cos \phi
      \left (2 \, \sqrt {E}
       + V_ 0\sqrt {\sin^2 \phi + 1} \right)
    - i\, V_ 0 \sin^2 \phi \right ]} \,.
\end{align}

Because the evanescent-wave solutions decay off as we move away from the delta-function potential (at $x =0$), they do not contribute to the reflection and transmission probabilities. Hence, the transmission and reflection coefficients, at an energy $E$, are given by 
\begin{align}
\label{eqT2D2}
& T ( E ,  V_0,\eta , \phi ) = | t ( E, V_0,\eta , \phi )|^2 \text{ and }
R ( E ,  V_0,\eta , \phi) = | r( E, V_0,\eta , \phi)|^2 ,\nn
& \text{with }
\phi = \tan^{-1} \left(  k_y/ \sqrt{E - k_y \, (k_y + \eta )} \right).
\end{align}
Here, $\phi  $ denotes the incident angle of the incoming plane wave. 
We represent the transmission-characteristics in the top and middle panels of Fig.~\ref{figtrs2D}, with the help of polar plots. While the top panel captures the untilted case (i.e., with $\eta =0 $), the middle panel corresponds to $\eta = 0.25$.
It is worthwhile to compare the results obtained above with those for a 2D electron-gas (since free electrons do have a quadratic-in-momentum dispersion). For a delta-function barrier, it is well-known that the transmission amplitude and the transmission probability for the free electrons are given by
\begin{align}
t (E, V_0, \phi )  =\frac { 2\, \sqrt {E} \, \cos \phi}
{ 2\,  \sqrt { E}\, \cos \phi + i\, V_0}
 = \frac {4 \, E}
{4\, E + V_ 0^2 \, \sec^2 \phi} \,.
\label{eqtval}
\end{align}
Comparing this with Eqs.~\eqref{eqtrs2D} and \eqref{eqtrs2D2}, we find that, while the free electrons' transmission probability is insensitive to the sign of $V_0$, the behaviour for 2D untilted QBCP changes as we switch the sign of $V_0$ [cf. Fig.~\ref{figtrs2D}]. The bottom panel of Fig.~\ref{figtrs2D} illustrates the polar plots of the transmission probability for a 2D electron-gas.

\textcolor{black}
{Let us discuss the features observed in Fig.~\ref{figtrs2D}:
\begin{enumerate}
\item As dictated by the form shown in Eq.~\eqref{eqtrs2D2}, there are $V_0$-dependent terms appearing in $T (E,V_0, \phi ) $, which make the transmission coefficients different for $\pm V_0$ when we consider 2D untilted QBCP. On the other hand, the free electrons' transmission coefficients are identical for $\pm V_0$ since the final expression is a function of $V_0^2 $ (and not $V_0$).
\item Both the untilted QBCP and an electron-gas have isotropic dispersions. Consequently, $T (E,V_0, \phi ) $ is a function of $\cos \phi $ for both cases. This feature is reflected in the polar plots with the curves being invariant under $\phi \rightarrow - \,\phi $. This situation gets altered as soon as we have a nonzero tilt in the spectrum, making it anisotropic with respect to the $k_y$-component of the momentum. This characteristic is observed in the middle panel.
\item The physically admissible values of the transverse momentum ($k_y $) lie in the range between $ - \,\sqrt{ E +\frac{\eta^2} {4}} - \frac{\eta}{2} $
and $\sqrt{ E +\frac{\eta^2} {4}} - \frac{\eta}{2} $ for a given value of incident energy ($E$).
Since $\phi = \tan^{-1} \left(  k_y/ \sqrt{E - k_y \, (k_y + \eta )} \right) $, the maximum value of $\phi$ is less than 90$^{\circ}$. Therefore, in the middle panel, many curves do not continue till 90$^{\circ}$.
\end{enumerate}
}

Let us assume $ W $ to be very large such that $k_y$ can effectively be treated as a continuous variable. Taking the continuum limit corresponding to $W \gg L$, we consider the zero-temperature limit. Remembering that $ \hbar =1 $ in the natural units, the conductance is given by [cf. Eq.~\eqref{eqland}]
\begin{align}
G (E,V_0) & 
 = \frac{e^2\, W} { 4\,\pi^2} 
\int_{- \,\sqrt{ E +\frac{\eta^2} {4}} - \frac{\eta}{2}}
^{\sqrt{ E +\frac{\eta^2} {4}} - \frac{\eta}{2}}
dk_y \, T(E, V_0,\eta, k_y)\,.
\end{align}
In the top and middle panels of Fig.~\ref{figfano2D}, we illustrate the conductance and
the Fano factors [see Eq.~\eqref{eqfano} for the formula of the Fano factor] for 2d QBCPs, as functions of $E$, for six different values of $ V_0$. Side by side, the corresponding curves [generated using Eq.~\eqref{eqtval}] for a 2D electron-gas are provided in the lowermost panel of Fig.~\ref{figfano2D}, for the sake of comparison.

\section{3D QBCP}
\label{sec3Dmodel}

A QBCP in a 3D system harbours a fourfold band-crossing, where the four bands form a four-dimensional representation of the underlying lattice-symmetry group \cite{MoonXuKimBalents}. The effective continuum Hamiltonian, derived using the $\left(\mathbf{k} \cdot \mathbf{p} \right)$ method, is written down using a convenient representation of the set of five $4\times 4$ Euclidean Dirac matrices, $\lbrace \Gamma_{a} \rbrace $, as shown below \cite{murakami, igor16}:
 \begin{align}
{\tilde{\mathcal{H}} }_{3D}^{\rm kin}(k_x,k_y,k_z) = \frac{1} {2\,m}
 \sum_{a=1}^5 d_a(\mathbf {k}) \,  \,\Gamma_a 
 + \tilde \eta \, k_y\, \mathbb{I}_{4 \times 4}  \,.
\label{eqbare}
 \end{align}
The $\Gamma_a$-matrices form one of the (two possible) irreducible four-dimensional Hermitian representations of the five-component Clifford algebra, defined by the anticommutator $\{ \,\Gamma_a, \,\Gamma_b \} = 2\, \delta_{ab}$.
The five anticommuting gamma-matrices can always be chosen such that three are real and two are imaginary \cite{murakami,igor12}. In the representation used here, $(\Gamma_1, \Gamma_3, \Gamma_5)$ are real and $(\Gamma_2, \Gamma_4 ) $ are imaginary:
\begin{align}
\Gamma_1 = \sigma_1 \otimes \sigma_0 \,, \quad  \Gamma_2 = \sigma_2 \otimes \sigma_0 \,,
\quad \Gamma_3 = \sigma_3 \otimes \sigma_1 \,, \quad \Gamma_4 = \sigma_3 \otimes \sigma_2 \,,\quad
\Gamma_5 = \sigma_3 \otimes \sigma_3 \,.
\end{align}
The five functions $  d_a(\mathbf{k})$ are the real $ \ell =2$ spherical harmonics (where $\ell$ represents the angular-momentum channel), with the following structure:
\begin{align}
& d_1(\mathbf k)= \sqrt{3}\,k_y\, k_z \, , \,\,  
d_2(\mathbf k)= \sqrt{3}\,k_x \,k_z \,, \,\, 
d_3(\mathbf k)= \sqrt{3}\,k_x\, k_y  , \quad
 d_4(\mathbf k) = \frac{\sqrt{3}}{2}(k_x^2 -k_y^2) \,, \,\, 
d_5(\mathbf k) =
\frac{1}{2}\left (2\, k_z^2 - k_x^2 -k_y^2 \right ) .
\end{align}
The second term in Eq.~\eqref{eqbare} represents a tilting of the dispersion with respect to the $k_y$-momentum direction.

Analogous to the 2D case, we multiply all the components of the momenta by $ \sqrt{2\,m}$. Hence, we deal with the scaled-momentum Hamiltonian captured by
\begin{align}
{\mathcal{H}}_{3D}^{\rm kin}(k_x,k_y,k_z) = 
 \sum_{a=1}^5 d_a(\mathbf {k}) \,  \,\Gamma_a 
 +  \eta \, k_y\, \mathbb{I}_{4 \times 4}  \,.
\label{eqham3D}
 \end{align}
The energy eigenvalues are doubly-degenerate with the values $\eta \, k_y \pm \, \varepsilon_{3D}(k_x,k_y,k_z)$, where
\begin{align}
\varepsilon_{3D}(k_x,k_y,k_z) = k_x^2 + k_y^2 +k_z^2  \, .
\end{align}
The ``$+$" and ``$-$" signs, as usual, refer to the conduction and valence bands, respectively.

\begin{figure*}[h!]
\subfigure{\includegraphics[width =  \textwidth]{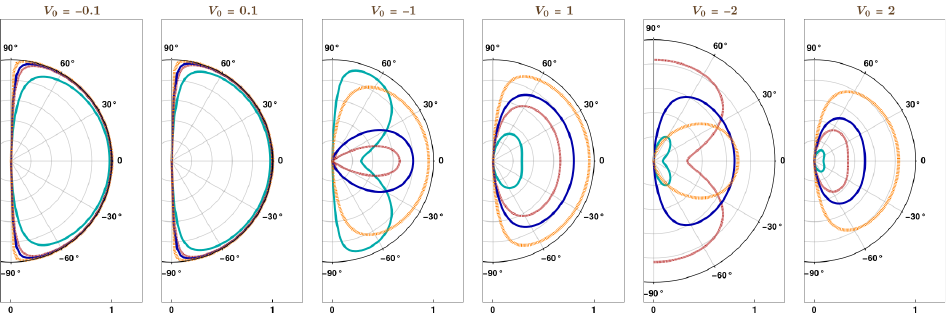}}
\subfigure{\includegraphics[width =  \textwidth]{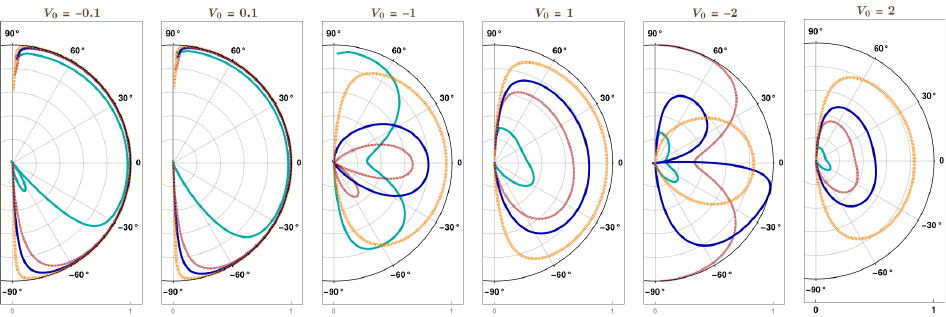}}
\subfigure{\includegraphics[width =  \textwidth]{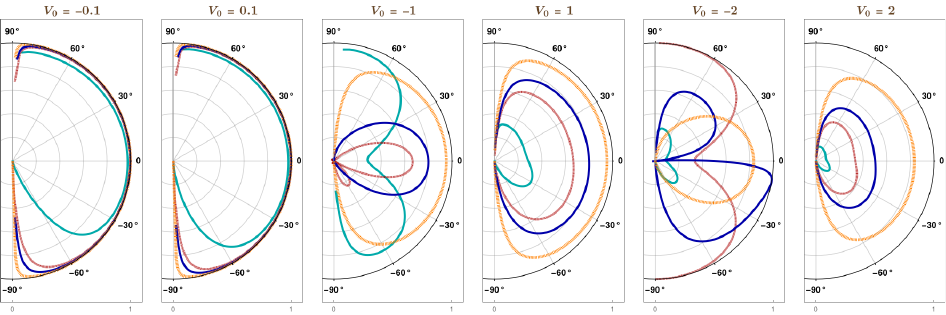}}
\subfigure{\includegraphics[width = 0.5 \textwidth]{legr}}
\caption{Plots of the transmission coefficient ($T$) as functions of the incident angle $\theta$, for various values of $ E $ and $V_0$. The subfigures in the top, middle, and bottom panels correspond to untilted 3D QBCP, tilted 3D QBCP with $\lbrace \eta, \phi \rbrace  = \lbrace 0.25 , \pi /2 \rbrace $, and tilted 3D QBCP with $\lbrace \eta, \phi \rbrace  = \lbrace 0.25 , 
\pi /4 \rbrace $, respectively.
\label{fig3Dtrs}}
\end{figure*}

A set of orthonormal eigenvectors is given by the following:
\begin{align}
\Psi_{-,1}^{T} & = \frac{1} { \mathcal{N}_{-,1} } \Big ( -
\frac{(k_{x}+ i\, k_{y}) \left (k+k_{z}\right )}
{(k_{x}- i\, k_{y})^{2}} \qquad
i\,  \frac{  k+3 \,k_{z}}
{\sqrt{3} \,(k_{x}- i\, k_{y})} \qquad -
i\, \frac{ -2\, k_{z} \left (k+k_{z}\right )+k_{x}^{2}+k_{y}^{2}}
{\sqrt{3}\, (k_{x}- i\, k_{y})^{2}} \qquad 1 \Big ),
\nonumber
\\
\Psi_{-,2}^{T} &= \frac{1} { \mathcal{N}_{-,2} } \Bigg (
\frac{(k_{x}+ i\, k_{y}) \left (k-k_{z}\right )}
{(k_{x}- i\, k_{y})^{2}} \qquad -
 i\, \frac{ \left (k-3\, k_{z}\right )}
{\sqrt{3} \,(k_{x}- i\, k_{y})} \qquad -
i\, \frac{ 2\, k_{z} \left ( k-k_{z}\right )+k_{x}^{2}+k_{y}^{2}}
{\sqrt{3}\, (k_{x}- i\, k_{y})^{2}} \qquad 1 \Bigg ) ,
\nonumber
\\
\Psi_{+,1}^{T} &= \frac{1} { \mathcal{N}_{+,1} } \Bigg( -
i\, \frac{ k+k_{z}} {\sqrt{3}\, (k_{x}- i\, k_{y})}
\qquad \frac{ k-k_{z}} {k_{x}+ i\, k_{y}} \qquad 1 \qquad -
i\,\frac{ 2\, k_{z} \left (k_{z}-k \right )+
k_{x}^{2}+k_{y}^{2}}
{\sqrt{3}\, (k_{x}+ i\, k_{y})^{2}} \Bigg ) ,
\nonumber
\\
\Psi_{+,2}^{T} &= \frac{1} { \mathcal{N}_{+,2} } \Bigg(
i\,\frac{k-k_{z}} {\sqrt{3}\, (k_{x}- i\, k_{y})}
\qquad -\frac{k+k_{z}}{k_{x}+ i\, k_{y}} \qquad 1 \qquad -
i\, \frac{2 \,k_{z}
\left (k+k_{z}\right )+k_{x}^{2}+k_{y}^{2}}
{\sqrt{3}\, (k_{x}+ i\, k_{y})^{2}} \Bigg ),
\label{eq23}
\end{align}
where $k= \sqrt{k_x^2  + k_y^2 +k_z^2}$. The ``$+$" (``$-$") subscript indicates an eigenvector corresponding to the positive (negative) eigenvalue. The symbols $\frac{1} { \mathcal{N}_{\pm,1}}$ and $\frac{1} { \mathcal{N}_{\pm,2}}$ denote the corresponding normalization factors.

The 3D system is modulated by a delta-function-like electric potential barrier/well,
\begin{align}
V ( z ) =  V_0 \,\delta(z)\,,
\label{eqpot2}
\end{align}
positioned at $z=0$. Here, we choose the $z$-axis as the transport direction, and place the chemical potential at an energy $E >0$ in the regions outside the potential barrier. Since the tilt-axis is $ k_y$, it does not matter whether we choose the propagation with respect to the $k_x$- or $k_z$-momentum --- our choice here is merely a matter of convenience. As before, we need to consider the total Hamiltonian in the position space as
\begin{align}
\mathcal{H}_{3D}^{\rm tot} &=
 \mathcal{H}_{3D}^{\rm kin}(- i\,\partial_x, - i\,\partial_y, - i\,\partial_z)+V(z) \,.
\end{align}

\subsection{Boundary conditions}

We consider the tunneling of the quasiparticles in a slab of height and width $W$ (i.e., with a square cross-section).
Again, we assume that the material has a sufficiently large width $W$ along each of the two transverse directions, such that the specific form(s) of the boundary conditions are irrelevant for the bulk response. Thus, for a wavefunction, $ \tilde \Psi ^{\mathrm{{tot}}}(x,y,z) $, in the position space, we employ the periodic boundary conditions obeying
$ \tilde \Psi ^{\mathrm{{tot}}}(x,0,z) = \tilde \Psi ^{\mathrm{{tot}}}(x, W,z)$ and
$\tilde \Psi ^{\mathrm{{tot}}}( 0,y,z) = \tilde \Psi ^{\mathrm{{tot}}}(W , y,z) $.
The transverse-momentum components, $ k_x $ and $ k_y$, are conserved, leading to modes with the dependence $ \sim e^{ i\,k_x\, x}\, e^{i\,k_y\, y} $. For any mode of a given set of values for the transverse-momentum components, we can determine the $z$-component of the wavevectors for the incoming, reflected, and transmitted waves (denoted by $k_{{\rm in}}$), by solving $
 E^2 = \left [ k_z^2 + k_x^2 + k_y \left( k_y +\eta \right )\right ]^2 \Rightarrow k_z^2
 = \pm \, E - k_\perp \left( k_\perp +\eta \sin \phi \right) $, where $k_{\perp } = \sqrt{k_x^2  +  k_y^2 }$, $k_x = k_\perp \cos \phi $, and $k_y = k_\perp \sin \phi $.
Hence, we obtain the following four solutions: 
\begin{align}
\label{eqksol2}
k_z = \pm \,\sqrt{  E -k_\perp \left( k_\perp +\eta \sin \phi \right)  } \text{ and }
k_z = \pm \, i\, \sqrt{ E +  k_\perp \left( k_\perp +\eta \sin \phi \right) } \,.
\end{align}
As we have understood by now from the 2D case, the imaginary solutions for $k_{z}$ imply the presence
of the evanescent waves.

\begin{figure}[t]
\subfigure{\includegraphics[width = 0.75 \textwidth]{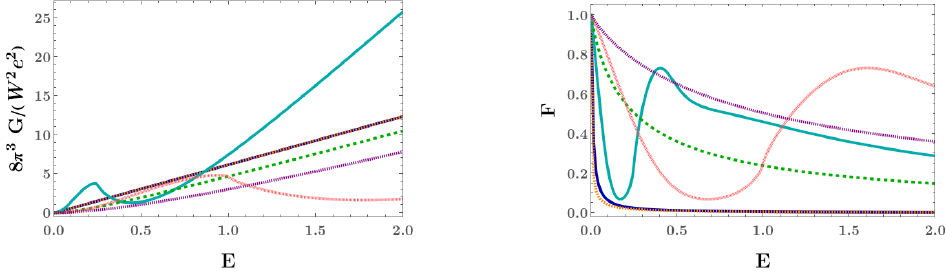}}
\subfigure{\includegraphics[width = 0.75 \textwidth]{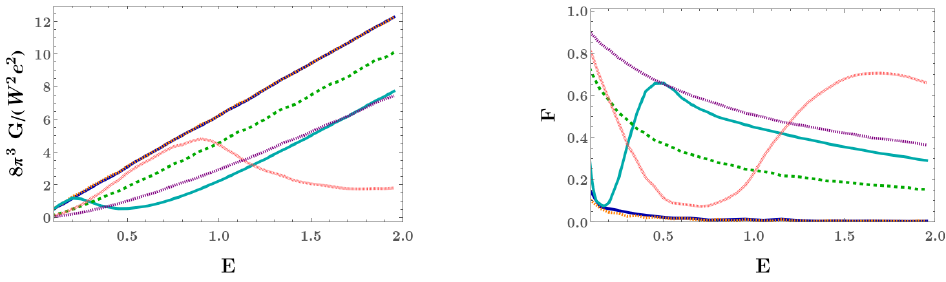}}
\subfigure{\includegraphics[width = 0.75 \textwidth]{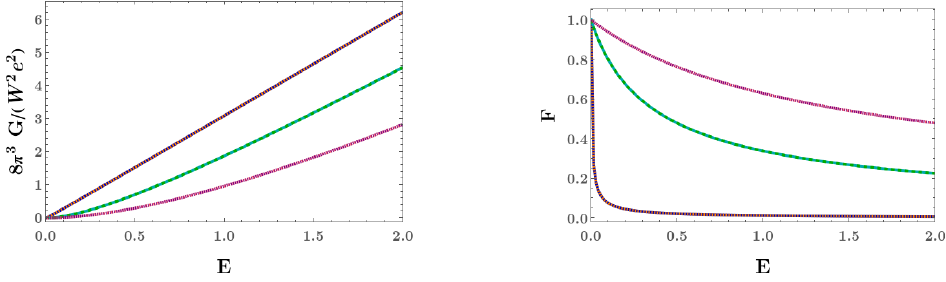}}
\subfigure{\includegraphics[width = 0.7 \textwidth]{leg}}
\caption{Plots of the conductance ($ G $) and Fano factor ($F$), as functions of $E $, for various values of $ V_0$. The subfigures in the top, middle, and lowest correspond to untilted 3D QBCP, tilted 3D QBCP with $\eta = 0.25 $, and 3D electron-gas, respectively.
\label{figfano3D}}
\end{figure}

To obtain the transport coefficients, we will follow the same procedure as described for the 2D QBCP. Without any loss of generality, we consider the transport of one of the degenerate positive-energy states, $\Psi_{+,1}$, corresponding to electron-like particles for one of the two degenerate conduction bands, with the Fermi level outside the potential barrier being adjusted to a value $E =\varepsilon_{3D} (k_x,k_y, k_z)  + \eta\,k_y $ (with $E>0$).
In this case, a mode $\sim  \tilde \Psi_{k_x, k_y} (z) \, e^{i \,k_x \, x}\, e^{i \,k_y \,y}$ is
constructed in a piecewise fashion as
\begin{align}
 \tilde \Psi_{ k_x,k_y} (z)=
  \begin{cases} \tilde \phi_L (z)  & \text{ for }  z  \leq 0   \\
\tilde \phi_R (z)  &  \text{ for } z > 0
\end{cases} ,
\end{align}
where
\begin{align}
 & \tilde \phi_L (z) = \frac{   
 \Psi_{+,1} ( k_x,k_y,  k_{\rm in}) 
 \, e^{ i\, k_{\rm in} \, z }
+  
\sum \limits_{s =1,2}
 r_{s} \,\Psi_{+,s} ( k_x,k_y, - \,k_{{\rm in}} ) 
 \, e^{- i\, k_{\rm in} \, z }
}
{\sqrt{ \tilde{ \mathcal{V} } }}
+ 
\sum \limits_{s =1,2}
{\tilde r}_{s} \,
\Psi_{-,s} (  k_x,k_y, -\, i \, \kappa ) 
 \, e^{ -\, \kappa  \, |z| }
\, ,\nn
& \tilde \phi_R (z) = 
\frac{ \sum \limits_{s =1,2}
 t_{s} \,\Psi_{+,s} ( k_x, k_y, k_{\rm in} ) 
 } 
{\sqrt{ \tilde{ \mathcal{V} } }}
\, e^{ i\, k_{\rm in} \, z}
+ 
\sum \limits_{s =1,2}
{\tilde t}_{s} \,
\Psi_{-,s} (  k_x, k_y, i \, \kappa ) 
 \, e^{   \kappa  \, z} ,\nn
& k_{{\rm in}}  = \sqrt{E - k_\perp \left( k_\perp +\eta \sin\phi \right)}\,, \quad
\tilde{\mathcal{V}} \equiv   
 |\partial_{k_{\rm in}} \varepsilon_{3D} 
 ( k_x, k_y, k_{\rm in} )|
=  2\, k_{\rm in} \,, \quad
\quad \kappa = \sqrt{  E + k_\perp \left( k_\perp +\eta \sin\phi \right) } \,.
\end{align}
Here, the magnitude of the group velocity, defined by $ \tilde{ \mathcal{V} } $, is needed to define the reflection and transmission amplitudes appearing in the unitary scattering matrix.

The boundary conditions can be obtained by integrating the equation, $\mathcal{H}_{3D}^{\rm{tot}}\, \tilde \Psi^{\rm{tot}} = E \, \tilde \Psi^{\rm{tot}} $, over a small interval around $ z =0$ in two consecutive steps. This amounts to imposing the continuity of the wavefunction and a constraint on its first-order derivatives (with respect to $z$), as shown below:
\begin{align}
\label{eqbdy2}
& \tilde \phi_L (0) =\tilde  \phi_R (0) \text{ and } 
\partial_z {\tilde  \phi_R}  (z) \big \vert_{z=0}
- \partial_z {\tilde \phi_L} (z)  \big \vert_{z=0} 
=   V_0 \, {\tilde  \phi_L } (0)\,. 
\end{align}
These conditions are sufficient to guarantee the continuity of the flux of the probability-current density along the $z$-direction. From the matching of the wavefunction and its derivatives, we have two matrix-equations from the two boundary conditions.
For 3D QPCBs, each of these matrix-equations can be separated into four components, since each wavevector has four components. Therefore, we have $  2\times  4 = 8 $ equations for the eight undetermined coefficients $\lbrace r_1 , \, r_2, \, \tilde r_1, \, \tilde r_1, \, t_1, \, t_2 , \, \tilde t_1, \, \tilde t_2 \rbrace $.

\subsection{Transmission coefficients, conductance, and Fano factors}

The reflection and transmission coefficients at an energy $E$ are given by 
\begin{align}
& R( E ,  V_0, \eta ,\theta, \phi  ) = | r_{1}( E, V_0 , \eta,\theta, \phi )|^2 
+ | r_{2}( E, V_0, \eta ,\theta, \phi )|^2 \text{ and }
T( E ,  V_0, \eta,\theta, \phi  ) = | t_{1}( E, V_0 , \eta,\theta, \phi )|^2  
+ | t_{2}( E, V_0, \eta ,\theta, \phi )|^2 \,,\nn 
& \text{with }
\theta = \tan^{-1} \left(  k_\perp / 
\sqrt{E - k_\perp \left (k_\perp  +\eta \sin \phi \right )} \right).
\end{align}
Here, $\theta $ defines the incident angle of the incoming wave. Although we have not provided the explicit expressions for $t_{1}$ and $t_{2}$, one can check that $| t_{2} |^2 = 0$ when the incident state is assumed to be $ \Psi_{+,1} $.\footnote{Similarly, $| t_{1} |^2 = 0$ when the incident state is assumed to be $ \Psi_{+,2} $.} Since the explicit analytical expressions for $t_{1}$, $t_{2}$, $ r_{1}$, and $ r_{2}$ are extremely long, we refrain from showing them here. Instead, we represent their characteristics via Fig.~\ref{fig3Dtrs}, with the help of polar plots representing $ T (E, V_0 , \eta ,\theta, \phi)$ as functions of $\theta $, picking some representative values of $V_0$, $E$, $\eta $, and $\phi$. In order to point out the effects of asymmetry introduced by tilt, we have plotted the curves by artificially extending the $\theta$-ranges to span over $\tan^{-1} \left( -\, k_\perp / 
\sqrt{E - k_\perp \left (k_\perp  -\eta \sin \phi \right )} \right)$ as well.

Let us discuss the features observed from the numerical results shown in Fig.~\ref{fig3Dtrs}:
\begin{enumerate}
\item $T $-curves for $\pm V_0$ in 3D QBCP show different characteristics. This reflects that there are $V_0$-dependent terms analogous to the 2D QBCP case.
\item An untilted QBCP has isotropic dispersion. Consequently, $T (E,V_0, \phi ) $ is a function of $\cos \theta $ for both cases. This feature is reflected in the polar plots with the curves being invariant under $\theta \rightarrow - \,\theta $. This situation gets altered as soon as we have a nonzero tilt in the spectrum, making it anisotropic with respect to the $k_y$-component of the momentum. This characteristic is observed in the middle panel. The effects of the tilt couple with the value of the azimuthal angle ($\phi$) as well.
\item The physically admissible values of the transverse momentum ($k_\perp $) lie in the range between $ 0 $
and $ \sqrt{ E +\frac{\eta^2 \sin^2 \phi} {4}} 
- \frac{\eta \sin \phi } {2} $ for a given set of values for $E$ and $\phi$.
Since $ \theta = \tan^{-1} \left(  k_\perp / 
\sqrt{E - k_\perp \left (k_\perp  +\eta \sin \phi \right )} \right) $, the maximum value of $\theta $ is less than 90$^{\circ}$. Therefore, in the middle panel, many curves do not continue till 90$^{\circ}$.
\end{enumerate}

Again, we assume $ W $ to be very large, such that $k_x$ and $ k_y$ can effectively be treated as continuous variables. In the zero-temperature limit, the conductance is given by [cf. Eq. \eqref{eqland}]
\begin{align}
G (E,V_0) & =
2\times \frac{e^2\, W^2} { 8\,\pi^3 } \int_0^{2\pi} d\phi
\int_{0}^{\sqrt{ E +\frac{\eta^2 \sin^2 \phi} {4}} 
- \frac{\eta \sin \phi } {2}}
dk_\perp \, k_\perp
\; T(E, V_0,\eta, k_\perp) \,.
\end{align}
In order to account for the twofold degeneracy (since we have two independent conduction bands), we have included an extra factor of two.
In Fig.~\ref{figfano3D}, we illustrate the conductance and the Fano factors [using Eq.~\eqref{eqfano}], as functions of $E$, for six distinct values of $V_0$. The corresponding curves [generated using Eq.~\eqref{eqtval}] for a 3D electron-gas are also provided for the sake of comparison.

\section{Bound states}
\label{secbs}

\begin{figure}[t]
\subfigure[]{\includegraphics[width = 0.32 \textwidth]{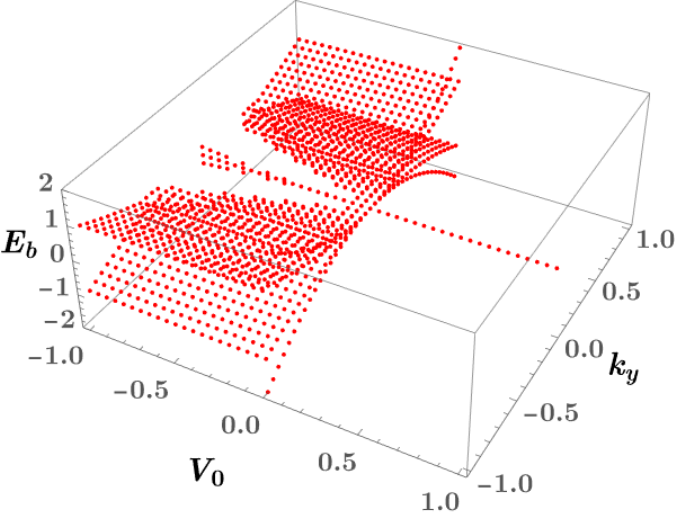}} \hspace{ 2 cm }
\subfigure[]{\includegraphics[width = 0.32 \textwidth]{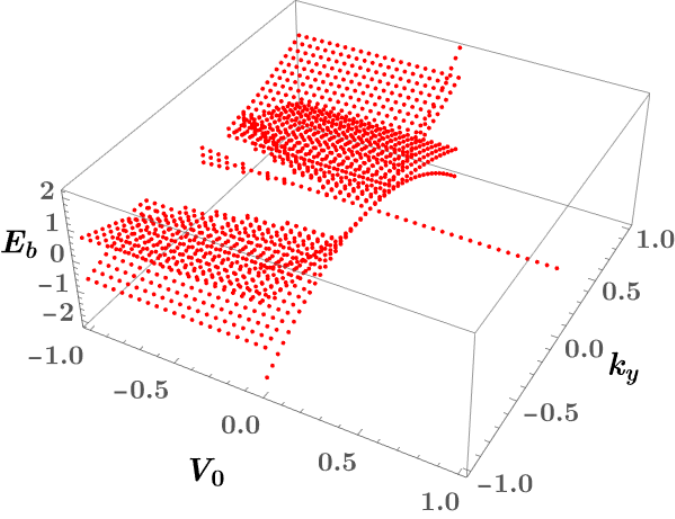}}
\subfigure[]{\includegraphics[width = 0.75 \textwidth]{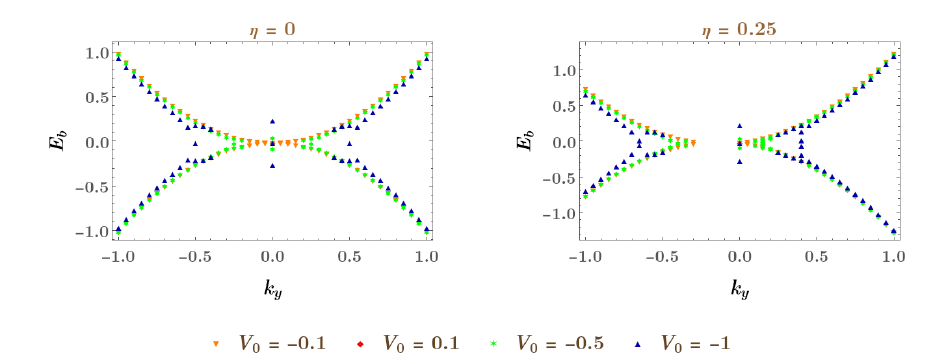}}
\caption{Energies of the bound states ($E_b$), as functions of $V_0 $ and $k_y$, for (a) an untilted 2D QBCP and (b) a tilted 2D QBCP with $\eta = 0.25 $. \textcolor{black}{In order to depict the asymmetry for nonzero $\eta$ more clearly, the values of $E_b$ have been shown in the subfigure (c) as functions of $k_y$ for some fixed values of $V_0$.}
\label{figbs2D}}
\end{figure}

In this section, we focus on determining the nature of the bound states appearing in our junction set-ups for QBCPs. We denote their energy-values by $ E_b $. We also discuss how the solutions differ from the solutions obtained for normal electron-gases.

\subsection{2D QBCP}

For the 2D case, we first define
\begin{align}
 & \phi^b_L (x)   =   r \, \Psi_\varsigma (  - \, k_{\rm in},  k_y) \, 
   e^{- i\, k_{\rm in} \, x }
+
\tilde r \, \Psi_{- \varsigma} (  - \,i \, \kappa ,  k_y) \,   e^{ - \kappa \,|x| } \, ,\quad
 \phi^b_R (x)  = 
  t\, \Psi_{\varsigma} ( k_{\rm in},  k_y)\,  e^{ i\, k_{\rm in} \, x }
+ \,\tilde t \, \Psi_{-\varsigma} ( i \, \kappa ,  k_y) \,   e^{- \,\kappa \, x } \,,\nn 
& k_{{\rm in}}  = \sqrt{  |E| - k_y \left(k_y+\eta \right) }\,, \quad
\kappa = \sqrt{ |E| + k_y \left(k_y+\eta \right)}\,, \quad \varsigma =\text{sgn} (E)\,.
\end{align}
Next, we need to impose the boundary conditions
\begin{align}
\label{eqbdy3}
& \phi^b_L (0) = \phi^b_R (0) \text{ and } 
\partial_x \phi^b_R  (x) \big \vert_{x=0}
- \partial_x \phi^b_L (x) \big \vert_{x=0} =   V_0 \, \phi^b_L (0)\,,
\end{align}
leading to $  2\times 2 = 4 $ linear homogeneous equations equations for the four undetermined variables, $\lbrace r, \, \tilde r,  \, t, \tilde t  \rbrace $.
Let $ \mathcal M_2$ denote the $ 4 \times 4  $ matrix constructed out of the four coefficients of these four variables. For the equations to be consistent, we need to impose the condition $\text{det}\, \mathcal M_2 = 0 $, whose solutions give us $E=E_b$. After some explicit calculations, we obtain
\begin{align}
\label{eqroots}
\text{det}\, \mathcal M_2 & = 
\left[k_y^2 \left (2 \, \kappa + V_ 0 \right)
   - \kappa \, k_ {\rm  in}
    \left (2 \, k_ {\rm  in} + i \, V_ 0 \right) \right]
\left[ V_ 0 \, k_y^2 - i \, k_ {\rm  in}
    \left \lbrace  2 \, k_y^2 + \kappa \left (2 \, \kappa +  V_ 0 \right) 
 \right  \rbrace \right ] .
\end{align}
In comparison, the bound-state energies for a 2D electron gas are given by $E_b = k_y^2 -  {V_ 0^2} / {4}$.
From Eq.~\eqref{eqroots}, we find that, for $k_y = 0 $, a 2D QBCP has $E_b \in \lbrace  0, \pm V_0^2 /4 \rbrace $. Therefore, there are two points to observe for $k_y=0$: (1) The bound states come with zero and positive values --- in fact, the nonzero values come in pairs of $\pm |E_b|$. (2) There exists a bound state with $E_b = 0 $ irrespective of the value of $V_0$. The numerically-evaluated values of $E_b$ are shown in Fig.~\ref{figbs2D}, where the appearance of paired values in the form of $ \pm |E_b| $ is observed. This is no surprise because the existence of the evanescent waves reflect the existence of bound states with positive value as well. A point to note is that nonzero $E_b $-values exist only for $V_0 < 0 $. For a nonzero $\eta$, an asymmetry with-respect-to the $k_y$-axis is observed, as expected from the tilting.

\begin{figure}[t]
\subfigure[]{\includegraphics[width = 0.32 \textwidth]{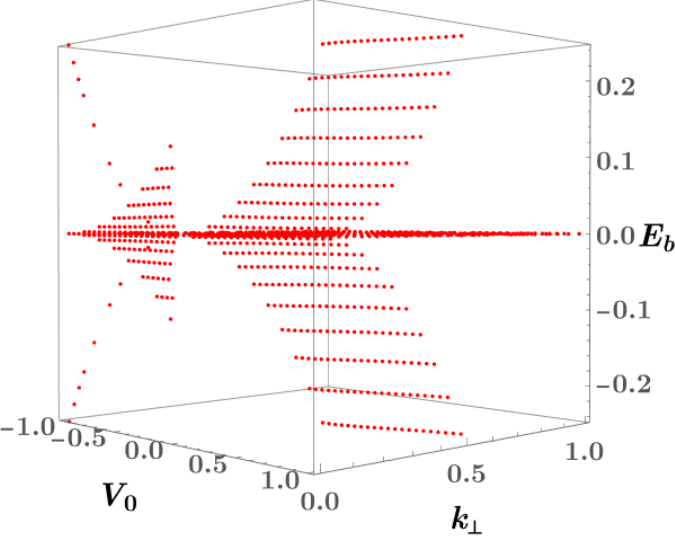}} 
\subfigure[]{\includegraphics[width = 0.32 \textwidth]{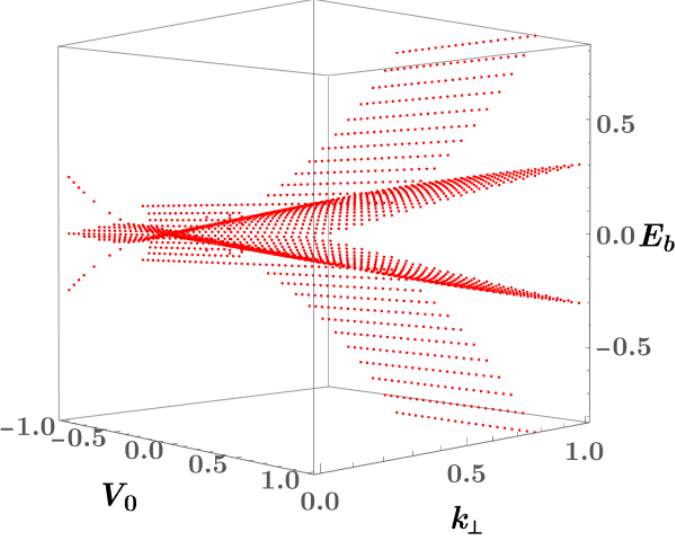}}
\subfigure[]{\includegraphics[width = 0.32 \textwidth]{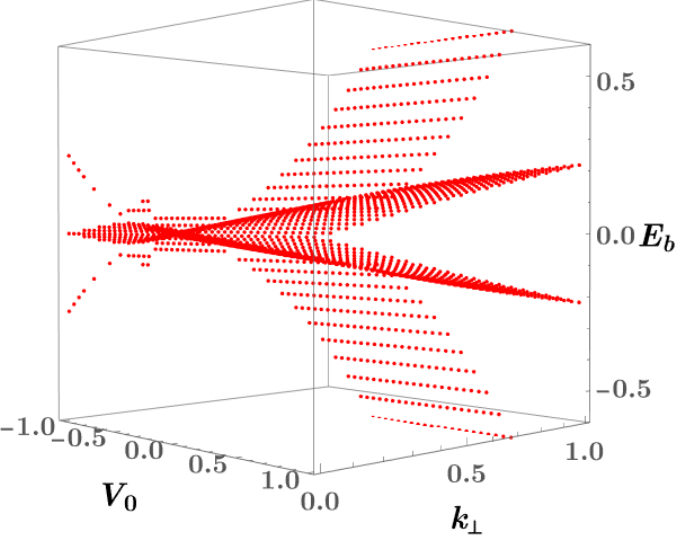}}
\subfigure[]{\includegraphics[width = 0.75 \textwidth]{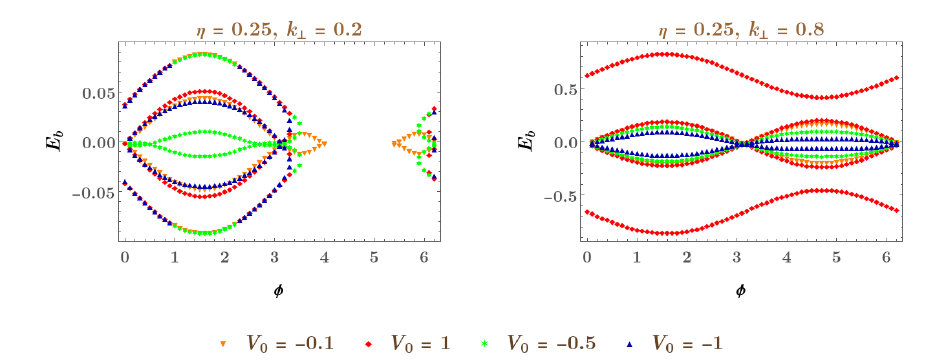}}
\caption{Energies of the bound states ($E_b$), as functions of $V_0 $ and $k_\perp$, for (a) an untilted 3D QBCP, (b) a tilted 3D QBCP with $\lbrace \eta, \phi \rbrace  = \lbrace 0.25 , \pi /2 \rbrace $, and (c) a tilted 3D QBCP with $\lbrace \eta, \phi \rbrace  = \lbrace 0.25 , \pi / 4 \rbrace $. \textcolor{black}{In order to depict the anisotropy caused in the $k_x k_y$-plane for nonzero $\eta$, the values of $E_b$ have been shown in the subfigure (c) as functions of $\phi $ for some fixed values of $V_0$ and $k_\perp$.}
\label{figbs3D}}
\end{figure}

\subsection{3D QBCP}

For the 3D case, we need to start with the definitions
\begin{align}
 & \tilde \phi^b_L (z) =  
\sum \limits_{s =1,2}
\left [  r_{s} \,\Psi_{\varsigma ,s} ( k_x,k_y, - \,k_{{\rm in}} ) 
 \, e^{- i\, k_{\rm in} \, z }
+
{\tilde r}_s \,
\Psi_{- \varsigma ,s} (  k_x,k_y, -\, i \, \kappa ) 
 \, e^{ -\, \kappa  \, |z| } \right ] ,\nn
& \tilde \phi^b_R (z) = 
 \sum \limits_{s =1,2} \left [ 
 t_s \,\Psi_{\varsigma  ,s} ( k_x, k_y, k_{\rm in} ) 
\, e^{ i\, k_{\rm in} \, z}
+ 
{\tilde t}_s \,
\Psi_{- \varsigma ,s} (  k_x, k_y, i \, \kappa ) 
 \, e^{    \kappa  \, z} \right ],\nn
& k_{{\rm in}}  = \sqrt{  |E| - k_\perp \left(k_\perp +\eta \sin \phi \right) }\,, \quad
\kappa = \sqrt{ |E| + k_\perp \left(k_\perp +\eta \right)}\,, 
\quad \varsigma =\text{sgn} (E)\,.
\end{align}
Next, the boundary conditions of
\begin{align}
\label{eqbdy3}
& \tilde \phi^b_L (0) = \tilde \phi^b_R (0) \text{ and } 
\partial_z \tilde \phi^b_R  (z) \big \vert_{z=0}
- \partial_z \tilde \phi^b_L (z) \big \vert_{z=0} =   
V_0 \,\tilde \phi^b_L (0)
\end{align}
provide us with $  2\times 4 = 8 $ equations for the eight undetermined variables, $ \lbrace r_1 , \, r_2, \, \tilde r_1, \, \tilde r_1, \, t_1, \, t_2 , \, \tilde t_1, \, \tilde t_2 \rbrace  $.
Let $ \mathcal M_3 $ denote the $ 8 \times 8  $ matrix constructed out of the coefficients of these eight variables. For the equations to be consistent, we need to impose the condition $\text{det}\, \mathcal M_3 = 0 $, whose solutions give us $E=E_b$. After some explicit calculations, we arrive at
\begin{align}
\label{eqroots2}
\text{det}\, \mathcal M_3 & = \Big[ 
- \, 2 \, \kappa \, k_{\rm in} \, k_\perp^2
\left (8 \, \kappa^2 + 7 \, \kappa \,  V_0 + 3 \, V_ 0^2 \right)
+2\, k_{\rm in}^3
\left \lbrace k_\perp^2
\left (8 \, \kappa + V_0 \right)
   - 4 \, \kappa^2\left (2 \, \kappa + V_ 0 \right) \right \rbrace
 \nn & \hspace{ 0.75 cm }  
+
i \, V_ 0 \, k_{\rm in}^2 
\left \lbrace k_\perp ^2\left (14 \, \kappa + V_ 0 \right)
   - 4 \, \kappa^2\left (2 \, \kappa + V_0 \right) \right \rbrace   
+ 8 \, k_{\rm in} \, k_\perp^4 \left (2\, \kappa + V_ 0 \right)
+i \, V_ 0\, k_\perp^2 \left (2 \, \kappa + V_ 0 \right)
  \left (4 \, k_\perp^2 - \kappa^2 \right) \Big ]^2 .
\end{align}
In comparison, the bound-state energies for a 3D electron-gas is given by $E_b = k_\perp^2 -  {V_ 0^2} /{4}$.
Due to the overall square on the right-hand side of Eq.~\eqref{eqroots2}, each $E_b$ appears as a twofold-degenerate root, which results from the doubly-degenerate energy-bands of the parent Hamiltonian. In addition, we observe that, for $k_\perp = 0 $, the bound states are given by $E_b \in \lbrace  0, \pm V_0^2 /4 \rbrace $, analogous to the 2D case. 
In general, the evanescent waves result in the solutions appearing as
$\pm E_b $ for generic values of $V_0$, $ k_\perp $, and $\phi$. This is reflected in Fig.~\ref{figbs3D}, where the values of $E_b$ are shown for some representative parameter-values. Contrasting with the 2D case, we find that nonzero values of $E_b$ appear even for $V_0 > 0 $. \textcolor{black}{An important point to note is that there is a $\phi $-dependence only for nonzero $\eta$. This is because the tilt introduces an anisotropy in the dispersion in the $k_x k_y$-plane (spanning the transverse directions). Fig.~\ref{figbs3D}(c) illustrates this $\phi $-dependence for some fixed values of $V_0$ and $k_\perp$.}

\section{Summary and outlook}
\label{secsum}

In this paper, we have elucidated the transmission characteristics of quasiparticles for generic tilted QBCPs, while travelling across a delta-function potential barrier/well. Compared to nodal-point semimetals harbouring linear-in-momentum band-crossings, the novelty of a parabolic spectrum is reflected by the necessity to include evanescent waves, while constructing the wavefunctions in a piecewise manner. The existence of these imaginary-wavevector solutions gives rise to bound states appearing as pairs of $\pm |E_b|$. We would like to emphasize that such exponentially-decaying solutions do not appear for the case of linear-in-momentum dispersive spectra, for example, which appear in Dirac cones of graphene \cite{geim, allain}, pseudospin-$1$ (or triple-point) semimetals \cite{fang, zhu, ips3by2}, Rarita-Schwinger-Weyl semimetals \cite{ips3by2}, and Weyl semimetals \cite{mansoor, deng2020, ips-aritra, ips-jns}. A nonzero tilt makes the situation even richer, with the tilt-anisotropy being reflected in anisotropies of the transmission coefficients, as well as the distribution of the bound states with respect to the values of the $k_y$-momentum. The twofold-degeneracy of the eigenvalues of the Hamiltonian in 3D QBCP results in the bound-state energies being doubly-degenerate as well.
Analogous systems with nonlinear-in-momentum dependence include semi-Dirac semimetals \cite{banerjee, ips-abs-semid} and multi-Weyl semimetals \cite{deng2020, ips-aritra, ips-jns}.

In the future, it will be worthwhile to investigate the nature of Andreev bound states, and their contributed currents in Josephson junctions, constructed out of the 2D and 3D QBCPs. This will be in the same spirit as has been studied for materials such as graphene, Weyl/multi-Weyl semimetals, semi-Dirac semimetals, pseudospin-1 semimetals, and Rarita-Schwinger-Weyl semimetals \cite{krish-moitri, debabrata-krish, ips-abs-semid, ips_jj_rsw, ips-jj-spin1}. However, the calculations will be significantly more challenging because of the proliferation of the number of undetermined coefficients, caused by the overarching presence of the evanescent waves.


\bibliography{biblio}

\end{document}